\documentclass[preprintnumbers,amsmath,amssymb]{revtex4}

\usepackage{graphicx}% Include figure files
\usepackage{amssymb}
\usepackage{bm}% bold math
\usepackage{color} 
\begin{document}

\title{Nonequilibrium entropic bounds for Darwinian replicators}

\author{Jordi Pi\~nero$^{1,2}\footnote{jordi.pinero@upf.edu}$ and Ricard Sol\'e$^{1,2,3}\footnote{ricard.sole@upf.edu}$}
\affiliation{$^{1}$ICREA-Complex Systems  Lab, Universitat Pompeu Fabra,   08003   Barcelona,   Spain}
\affiliation{$^{2}$Institut de Biologia Evolutiva (CSIC-UPF), Psg Maritim Barceloneta, 37, 08003 Barcelona, Spain}
\affiliation{$^{3}$Santa Fe Institute, 399 Hyde Park Road, Santa Fe NM 87501, USA}

\date{\today}

\begin{abstract}
Life evolved on our planet by means of a combination of Darwinian selection and innovations leading to higher levels of complexity. The emergence and selection of replicating entities is a central problem in prebiotic evolution. Theoretical models have shown how populations of different types of replicating entities exclude or coexist with other classes of replicators. Models are typically kinetic, based on standard replicator equations. On the other hand, the presence of thermodynamical constrains for these systems remain an open question. This is largely due to the lack of a general theory of out of statistical methods for systems far from equilibrium. Nonetheless, a first approach to this problem has been put forward in a series of novel developements in non-equilibrium physics, under the rubric of the extended second law of thermodynamics. The work presented here is twofold: firstly, we review this theoretical framework and provide a brief description of the three fundamental replicator types in prebiotic evolution: parabolic, malthusian and hyperbolic. Finally, we employ these previously mentioned techinques to explore how replicators are constrained by thermodynamics.
\end{abstract}

\keywords{Evolution; Replicators; Life; Entropy; Thermodynamics}

\maketitle

\section{Introduction}
Biology follows the laws of physics, and yet it remains distinctive from many standard physical systems in a number of ways. In the first place, life's self-replicating mechanisms stand as a major difficulty when approaching it as a simple physical setup. On the other hand, life too differs from physics in its computational nature: all living forms conduct some sort of computation as a crucial component of their adaptive potential \cite{Hopfield94-JTB}. The success of life over chemistry is largely associated to the emergence of prebiotic molecular mechanisms that, in turn, allowed for a template-based landscape to become dominant over the whole biosphere. How this took place is one of the most fundamental questions in science \cite{Smith99-OriginsLife, Dyson99-OriginsOfLife, Kauffman93-OriginsOrder}.  

Life forms are out-of-equilibrium structures able to employ available matter, energy and information to propagate some type of identity. Most theoretical approaches to the evolution of replicators have been grounded on a kinetic description. Under such framework, interactions between (typically molecular) agents are represented by nonlinear differential equations, known as {\em replicator} equations \cite{Hofbauer98-EvoGames}. They provide a deterministic view of Darwinian dynamics. However, as pointed out by Smith and Morowitz, "the abstraction of the replicator, which reduces Darwinian dynamics to its essentials, also de-emphasizes the chemical nature of life" \cite{Morowitz16-OriginLifeEarth}. The same can be concluded in relation with the lack of a thermodynamical context. Despite early efforts towards the development of a physics of evolutionary dynamics \cite{Morowitz16-OriginLifeEarth,Babloyantz86-MoleculesDynamicsLife, Kauffman00-Investigations, Nicolis-ExploringComplex} a more satisfactory formalism has yet to emerge. In particular, life propagation processes require an entropy production and balance equations can be defined \cite{Nicolis-ExploringComplex, Glansdorff-ThermoStructure, Wagensberg00-BioPhi}. However, a more general non-equilibrium statistical physics approach suitable for the problem of self-replication has been missing until recently \cite{Jarzynski97-PRL, Crooks98-JStatPhys, Crooks99-PRE, Gomez-Marin08-PRE, Still12-PRL, England13-JChemPhys, Perunov16-PRX, Barlotta16-PRE}. How can this novel approach apply to the fundamental problem of replicator dynamics in the eary stages of Life on Earth?

Beyond the self-replicating potential of cells and molecules, several replication strategies are at work in living systems, also involving multiple scales \cite{Szathmary-JTheorBiol, Szathmary-PhilTrans, Sole16-PhilTrans}. The basic growth dynamics followed by each class has remarkably different consequences for selection. The simplest class is the Mathusian (exponential) growth dynamics exhibited by cellular systems growing under unlimited resources. Two other types of replicators are observed in Nature. One is associated to the emergence of cooperation dynamics, with different classes of replicators helping each other and forming a mutualistic assembly \cite{Eigen79-Hypercycle}. The second is related to a template-based replication mechanism that we can identify in living systems as the standard mechanism of nucleic acid replication. This mechanism has been shown to lead to the "survival of everyone": it provides a mechanism capable of sustaining very diverse populations of replicators \cite{ Szathmary89-JTB, Scheuring01-JTB}.

From the physics perspective, these systems involve large number of internal degrees of freedom interacting in an out-of-equilibrium context. Therefore, a thermodynamical framework explaining the qualitative differences exhibited by these classes is needed. The work presented here is an attempt to delineate the fundamental thermodynamical constrains for the three elemental types of prebiotic replicators.
 
%%%%%%%%%%%%%%%%%%%%%%%%%%%%%%%%%%%%%%%%%%

\section{Entropic Bounds for Replicators}

%%%%%%%%%%%%%%%%%%%%%%%%%%%%%%%%%%%%%%%%%%

Let us begin by reviewing the theoretical framework upon which the analysis of the problem will unfold \cite{Gomez-Marin08-PRE, England13-JChemPhys, Perunov16-PRX, Barlotta16-PRE}. Then, we summarize the elemental classes of replicators and their essential aspects \cite{Szathmary-JTheorBiol}, together with a series of implications regarding selection and adaptation. Finally, we lay out an approach to the question of how non-equilibrium thermodynamical bounds arise in these types of systems and how such constrains might have affected early evolution scenarios.

%%%%%%%%%%%%%%%%%%%%%%%%%%%%%%%%%%%%%%%%%%

\subsection{The Extenended Second Law}

Consider a classical time-evolving system described by its microscopical trajectory in the phase space $x(t)\in\Omega$ plus a set of controlled parameters $\lambda (t)$ both evolving in a time interval $t\in[0,\tau]$. Assume that the system remains in contact with a {\em heat bath} at temperature $T={ 1 /\beta }$ throughout the entire trajectory. Denote the transition probability from a miscroscopical state $x$ to $y$ in the time interval $\epsilon$ by $\pi_{\epsilon}[x\to y]$. Now, if we slice time as $t_{i+1}-t_i=\epsilon$, with $t_n=\tau=n\epsilon$ and $t_0=0$, then, for sufficiently small $\epsilon$, the {\em microscopical reversibility} condition implies \cite{Crooks98-JStatPhys, Crooks99-PRE}:

\begin{eqnarray}\label{Generalized Detailed Balance I}
\frac{\pi_{\epsilon}[x^*(\tau-t)]}{\pi_{\epsilon}[x(t+t_{n-1})]}\cdots\frac{\pi_{\epsilon}[x^*(t_1-t)]}{\pi_{\epsilon}[x(t)]}=\exp\left \{ -\beta \sum_{i=0}^{n-1}Q^{b}_{i\to i+1} \right \} \ ,
\end{eqnarray}
where the superscript $*$ denotes momentum-reversed microstates, and $Q^{b}_{i\to i+1}$ denotes the heat exchange in going from from states $x(t_i)$ to $x(t_{i+1})$ as measured from the heat bath. Heuristically, \eqref{Generalized Detailed Balance I} is interpreted as the composed detailed balance condition on each time-slice of the trajectory $x(t)$ (see Figure 1a). This can be represented by the functional relation:

\begin{eqnarray}\label{Generalized Detailed Balance II}
\frac{\pi_{\tau}[x^*(\tau-t)]}{\pi_{\tau}[x(t)]}=\exp\left \{ -\beta Q_{b}[x(t)] \right \} \ .
\end{eqnarray}

\begin{figure}[th]
\centering
\includegraphics[width=14 cm]{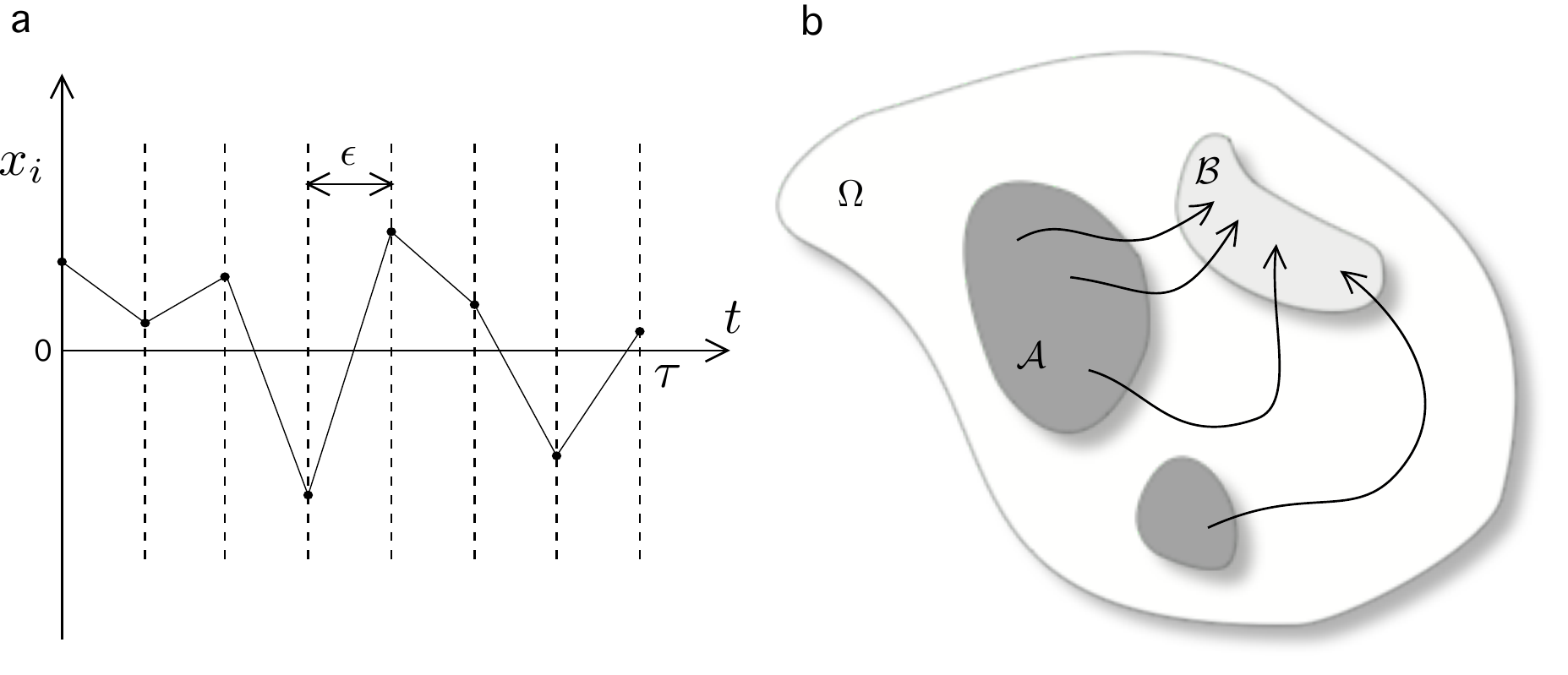}
\caption{Scheme of the formal approach to expressions \eqref{Generalized Detailed Balance I}-\eqref{Macroscopic Transition Rate II}. (\textbf{a}) A time-discretization is implemented in order to characterize the microscopical reversibility condition. (\textbf{b}) A qualitative scheme of possible trajectories between macrostates on the global phase space. The macroscopic coarse-grained states, $\mathcal{A}$ (dark shaded region) and $\mathcal{B}$ (light shaded region) are defined as disjoint ($\mathcal{A}\cup \mathcal{B} = \emptyset $) sections on the phase state $\Omega$. The set of forward paths of duration $\tau$ constrained to start in $\mathcal{A}$ and finish in $\mathcal{B}$ is denoted by ${\bf{x}}_{\tau}$.}
\end{figure}   

Next, let us introduce two {\em macrostates} which can be interpreted as two disjoint sections of the phase space, $\mathcal{A},\mathcal{B} \subset\Omega$ (see Figure 1b). Let us introduce notation for macrostate bounded trajectories in $\Omega$ by defining the set of forward trajectories ${\bf{x}}_{\tau}=\{x(t), \ t\in [0,\tau] \ | \  x(0)\in \mathcal{A} \ \wedge \ x(\tau)\in \mathcal{B} \}$ , i.e. the set of possible trajectories subject to condition that the initial microstate is in $\mathcal{A}$ and the final must be in $\mathcal{B}$. Then, construct the formal coarse-grained transition rate from $\mathcal{A}$ to $\mathcal{B}$ as

\begin{eqnarray}\label{Macroscopic Transition Rate I}
\Pi_{\tau}(\mathcal{A}\to \mathcal{B})=\int_{{\bf{x}}_{\tau}} \mathcal{D}[x(t)]\pi_{\tau}[x(t)] \ ,
\end{eqnarray}

while, equivalently, denote ${\bf{x}}^*_{\tau}=\{x^*(\tau-t), \ t\in [0,\tau] \ | \  x^*(\tau)\in B \ \wedge \ x^*(0)\in A \}$ , i.e. the set of reversed macrostate bounded trajectories, and compute the inverse coarse-grained transition rate from $\mathcal{B}$ to $\mathcal{A}$ as

\begin{eqnarray}\label{Macroscopic Transition Rate II}
\Pi_{\tau}(\mathcal{B}\to \mathcal{A})=\int_{{\bf{x}}^*_{\tau}} \mathcal{D}[x^*(\tau - t)]\pi_{\tau}[x^*(\tau -t)] \ .
\end{eqnarray}

Here onwards, let use bracket notation $\big{<} \cdot \big{>}$ to denote averages over forward paths ${\bf{x}}_{\tau}$. Under this theoretical framework, it can be shown \cite{England13-JChemPhys, Barlotta16-PRE} that the following relation must hold:

\begin{eqnarray}\label{Extended Jarzynski Equality I}
\left < \exp \left \{ - \Delta H [x(t)] - \beta Q_{b}[x(t)] + \log \left [ \frac{\Pi_{\tau}(\mathcal{A}\to \mathcal{B})}{\Pi_{\tau}(\mathcal{B}\to \mathcal{A})} \right ]\right \} \right > = 1\ ,
\end{eqnarray}

where we have defined the path dependant observable:

\begin{eqnarray}\label{Nonequilibrium Entropy Measure}
\Delta H [x(t)] = - \log \left [ \frac{p_{\tau} \left ( x(\tau) \right )}{p_0 \left ( x(0) \right )} \right ] \ ,
\end{eqnarray}

with $p_{\tau} \left ( x(\tau) \right )$ and $p_0 \left ( x(0) \right )$ standing for the probability of landing at a certain $x(\tau)\in\mathcal{B}$ at time $t=\tau$ and departing from $x(0)\in\mathcal{A}$ at time $t=0$. Notice that \eqref{Nonequilibrium Entropy Measure} is a functional on the boundary conditions of the trajectory $x(t)$. Let us define,

\begin{eqnarray}
\beta \mathcal{W}[x(t)]\equiv\Delta H [x(t)] + \beta Q_b[x(t)] \ ,
\end{eqnarray}

as a functional observable over the sample of forward paths ${\bf{x}}_{\tau}$. On the one hand, a first order expansion on \eqref{Extended Jarzynski Equality I} imposes the following boundaries to the fraction of the coarse-grained transition rates:

\begin{eqnarray}\label{England Bounds}
\log  \left [ \frac{\Pi_{\tau}(\mathcal{A}\to \mathcal{B})}{\Pi_{\tau}(\mathcal{B}\to \mathcal{A})} \right ]\leq \beta \left < \mathcal{W}[x(t)] \right > = \left < \Delta H [x(t)] \right > + \beta \left < Q_b[x(t)] \right > \ .
\end{eqnarray}

This results implicitely allude to the Landauer bounds on heat production for bit erasure \cite{Landauer-IBM, Bennett-IntJourTheorPhys, Parrondo-Nature}. Inequality \eqref{England Bounds} constrains the irreversibility of the macroscopic process $\mathcal{A}\to \mathcal{B}$ with respect to the average entropy produced internally, $\Delta H$, and externally (into the bath), $\beta Q_b$, and it is dubbed the {\em Extended/Bayesian Second Law} (ESL) \cite{England13-JChemPhys, Barlotta16-PRE}. One interpretation is that macroscopic irreversibility increases the minimum dissipated energy during the process $\mathcal{A}\to \mathcal{B}$. Interestingly, expression \eqref{England Bounds} formalizes a bound on entropy production defined only by the coarse-grained properties of such a process, i.e., dependent only on the macroscipic transition rates which, under certain circumstances, may be the only measurable quantities of a given system. We will come back to this point in the following sections.

On the other hand, a general perturbative analysis using the cumulant expansion \cite{Zwanzig57-JChemPhys} onto \eqref{Extended Jarzynski Equality I} leads to

\begin{eqnarray}\label{Extended Jarzynski Equality II}
\log  \left [ \frac{\Pi_{\tau}(\mathcal{A}\to \mathcal{B})}{\Pi_{\tau}(\mathcal{B}\to \mathcal{A})} \right ]=\sum_{l\geq 1} (-\beta)^{l-1}\frac{\omega_l}{l!}\ ,
\end{eqnarray}
where $\omega_l$ stands for the $l-$th cumulant of the distribution of $\beta \mathcal{W}[x(t)]$. In fact, \eqref{Extended Jarzynski Equality II} allows for a more sophisticated view of 

\begin{eqnarray}\label{Extended Jarzynski Equality III}
\log  \left [ \frac{\Pi_{\tau}(\mathcal{A}\to \mathcal{B})}{\Pi_{\tau}(\mathcal{B}\to \mathcal{A})} \right ]=\beta \left < \mathcal{W}\right >  - \Phi_{\tau}(\beta)\ ,
\end{eqnarray}
where, formally
\begin{equation}
\Phi_{\tau}(\beta)=\frac{\beta^2}{2}\left < \mathcal{W} \right >^2_c-\frac{\beta^3}{6}\left < \mathcal{W} \right >^3_c + \cdots
\end{equation}
with the subscript $c$ indicating cumulant expressions. Combining equations \eqref{Extended Jarzynski Equality I} and \eqref{Extended Jarzynski Equality II}, it can be shown that $\Phi_{\tau}\geq 0$. Indeed, $\Phi_{\tau}$ is a measure the fluctuations of the distribution associated to observable $\mathcal{W}[x(t)]$. Thus, equation \eqref{Extended Jarzynski Equality III} represents an extended fluctuation-dissipation theorem, where the LHS reflects the macroscopic (coarse-grained) irreversibility property and the RHS a balance between dissipated work and fluctuations over the ${\bf{x}}_{\tau}$ sample.

This result is of particular interest when a system is arranged such that a choice between two macroscopical end-states is forced. In such cases, fluctuation discrepancies might break symmetry thus favoring certain macroscopical transitions or supressing others \cite{Perunov16-PRX}.

In the following sections we will revisit the paradigm of prebiotic replicators, and focus on how to minimally embed this problem into the formalism discussed above. Subsequently, we will argue how these entropic constrains may have coupled to prebiotic selection and added preassure to in an evolutionary context.

\begin{figure}[t]
\centering
\includegraphics[width=15 cm]{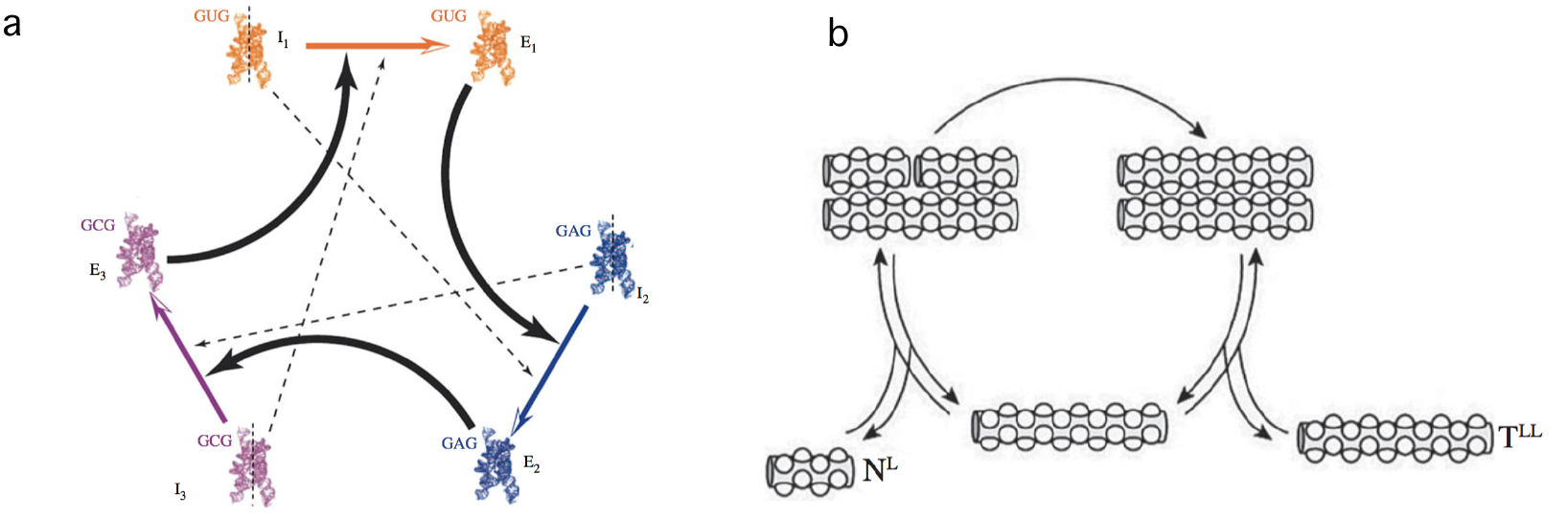}
\caption{Hyperbolic and parabolic replicators. In (a) we display a simplified scheme of an experimental implementation of a catalytic set of ribozymes forming a cooperative loop. Here 
each component of the system helps the next to replicate. Dashed lines indicate weaker catalytic 
links (modified from \cite{Vaidya12-Nature}). The parabolic system outlined in (b) 
is based on complementary (template) peptide chains involving a ligation mechanism (adapted 
from \cite{Lee96-Nature}). }
\end{figure} 

\subsection{Replicators \& Reproducers}

%%%%%%%%%%%%%%%%%%%%%%%%%%%%%%%%%%%%%%%%%%

Several fundamental replication strategies are at play in living systems. These strategies are present in multiple scales, from molecular replicators to cells and beyond. Each class of replicating agent is characterized by a kinetic pattern, which dynamics entail distinct selective implications. Here, we will focus on three characteristical replicator classes \cite{Szathmary-JTheorBiol, Szathmary-PhilTrans}.

{\em Simple} replicators: commonly known as Malthusian agents, correspond to systems whereby a single component $A$ is capable of making a copy of itself by using the available resources, namely $E$, generating a certain waste product, $W$. Schematically,

\begin{equation}\label{Simple Reaction}
A + E \stackrel{g}{\longrightarrow} A + A + (W) \ .
\end{equation} 

Assuming a large repository of resources, the kinetics of this process can be reduced to a linear dynamical equation (see Table 1). Systems following this mechanism obey exponential growth laws. 

{\em Hyperbolic} replicators: tied to one of the most relevant novelties in evolution \cite{Smith95-MajorTransitions, Szathmary91-TREE} is the concept of {\em autocatalysis}. This mechanism is a precursor of self-replicating entities that largely define the nature of living structures. It has been put forward by several authors \cite{Eigen71-NaturwissenchaftenI, Eigen77-NaturwissenchaftenII, Eigen77-NaturwissenchaftenIII, Eigen77-NaturwissenchaftenIV} as an central process in the chemistry of prebiotic systems involving the emergence of cooperative agents (see Figure 2a).

\begin{equation}\label{Hyperbolic Reaction}
A + A + E \stackrel{h}{\longrightarrow} A + A + A + (W) \ .
\end{equation} 

Again, under well-mixed and unlimited resource conditions, the hyperbolic replicator kinetics is reduced to a second order equation (see Table 1). Autocatalytic growth is characterized by displaying a finite-time singularity at $t_c=1/hx_0$ \cite{Szathmary-JTheorBiol}.

{\em Parabolic} replicators: this type of replicator arises from a combination of molecular reactions. In particular, oligonucleotides are known to exhibit such behaviour \cite{Kiedrowski-AngewChim, Zielinski-Nature, Paul04-CurrentOpinion}. The minimal scheme where this particular dynamics is observed consists of the set of processes (see Figure 2b).

\begin{equation}\label{Parabolic Reactions}
A + E \stackrel{c}{\longrightarrow} AA + (W) \overset{a}{\underset{b}{\leftrightarrows}} A + A + (W^{\prime}) \ ,
\end{equation} 

which, under conditions $a \gg b \gg c$ is reduced to a parabolic law $\dot{x}=\rho\sqrt{x}$, where $x$ denotes the total concentration of the molecular component $A$ regardless of the configuration, it being either associated ($AA$) or dissociated ($A$) (see Appendix A). Parameter $\rho=c\sqrt{2b / a}$.

\begin{table}[t]
\centering

\begin{tabular}{|c|c|c|}
\hline
\; \textbf{Replicator Class}\;\;\;\; & \;\;\;\;  \textbf{Reaction Scheme}\;\;\;\; 	& \;\;\;\;  \textbf{Effective Dynamics} \;\;\;\; \\
\hline

Simple		& $A + E \to A + A$ 			& $\dot{x}=gx$\\
Hyperbolic		& $A + A + E \to A + A + A$			& $\dot{x}=hx^2$\\
Parabolic		& $A + E \to AA \leftrightarrow A + A$ 			& $\dot{x}=\rho x^{1/2}$\\
\hline

\end{tabular}
\caption{Summary of the minimal expressions for the kinetics of the three replicator classes discussed above. We have denoted as $x$ the gross concentration of replicating molecules $A$, independently of the configuration.}
\end{table}

\subsection{Coarse-grained Dynamics of Replicators}

The dynamics of the three types of replicators discussed above are recognizably taking place on the macroscopic level. Molecular replicators encapsulate a whole system rich in complexity and structure, thus the measurable transition rates, such as $g$, $h$ or $\rho$ above, are merely an emergent feature of the interplay of the many internal degrees of freedom of the system.

Suppose that a system is composed of a fixed number of molecular templates or {\em chains}, $N$, which can either be internally ordered such that they behave as a replicators ($A$), namely {\em active chains} or simply act as substrate ($E$), namely {\em inactive chains}. The goal here is to define an unambiguous coarse-graining measure capable of distinguishing two meaningful macroscopic states of the system. To do so, we will consider three such systems which replicators' act accordingly with the three replicator classes summarized in Table 1. 

Considering a markovian approach \cite{Mendez14-Stochastic, Redner01-FirstPassage}, 
each set of reaction rules allows defining transition probabilities and a master equation that 
in general will read: 
\begin{eqnarray}\label{MEQ General}
\frac{dP(n,t)}{dt}= \sum_{m\neq n} \omega\left( n| m \right) P(m,t) - \sum_{m\neq n} \omega \left ( m | n \right) P(n,t) \ , 
\end{eqnarray}
and gives the probability $P(n,t)$ of observing $n$ active chains at time $t$. Here the 
$\omega(i\vert j)$ terms introduce the transition probabilities associated to each rule.The corresponding 
master questions associated to the Malthusian, hyperbolic and parabolic cases, respectively (see Appendix B for details):

\begin{eqnarray}\label{MEQ Simple I}
\frac{dP(n,t)}{dt} &=& g\left ( \frac{n}{N} \right ) \left ( 1 - \frac{n}{N} \right ) \left [ P(n-1,t)-P(n,t) \right ]  - \delta \left ( \frac{n}{N} \right ) \left [ P(n,t)-P(n+1,t) \right ]  \\
\label{MEQ Hyperbolic I}
\frac{dP(n,t)}{dt} &=& h\left ( \frac{n}{N} \right )^ 2 \left ( 1 - \frac{n}{N} \right ) \left [ P(n-1,t)-P(n,t) \right ]  - \delta \left ( \frac{n}{N} \right ) \left [ P(n,t)-P(n+1,t) \right ]  \\
\label{MEQ Parabolic I}
\frac{dP(n,t)}{dt} &=& \frac{bc}{2a} \left( \sqrt{1+\frac{4an}{bN}} - 1  \right ) \left ( 1 - \frac{n}{N} \right ) \left [ P(n-1,t)-P(n,t) \right ] \nonumber \\
&-& \delta \left ( \frac{n}{N} \right ) \left [ P(n,t)-P(n+1,t) \right ] 
\end{eqnarray}

Notice that \eqref{MEQ Simple I}, \eqref{MEQ Hyperbolic I} and \eqref{MEQ Parabolic I} are all macroscopic representations of the replicating dynamics. Here, the internal interactions that produce the effective behaviour described by the previous set of equations are all integrated out into its corresponding coupling constants. Thus, within this macroscopical framework we shall define the phase space subsets:

\begin{itemize}
\item	$\mathcal{A}$ - state in which the system contains a total number of $n-1$ active chains.
\item	$\mathcal{B}$ - state in which the system contains a total amount of $n$ active chains.
\end{itemize}

Let us focus on the explicit bounds given by the LHS in expression \eqref{Extended Jarzynski Equality I}, we first introduce notation for these {\em lower entropic bounds} as

\begin{eqnarray}\label{LEB Notation}
LEB_r(x):=\log \left [ \frac{\Pi_{\tau}(\mathcal{A}\to \mathcal{B})}{\Pi_{\tau}(\mathcal{B}\to \mathcal{A})} \right ] \ ,
\end{eqnarray}
where the subscript $r\in \{{\bf s},{\bf h},{\bf p}\}$ indicates the replicator type (simple, hyperbolic and parabolic respectively), while $x:=n/N$ in each case. Therefore, considering that the transition rates $\Pi_{\tau}(\mathcal{A}\to \mathcal{B})$ and $\Pi_{\tau}(\mathcal{B}\to \mathcal{A})$ for the defined coarse-grained states $\mathcal{A}$ and $\mathcal{B}$ correspond to the prefactors in each master equation above,

\begin{eqnarray}\label{LEB Simple Hyperbolic and Parabolic}
LEB_{{\bf s}}(x)=\log \left[ \frac{g}{\delta} (1-x) \right ] \ , \ \  LEB_{{\bf h}}(x)=\log \left[  \frac{h}{\delta} x(1-x)  \right ] \ , && \\
\label{LEB Parabolic}
LEB_{{\bf p}}(x)=\log \left [ \frac{c}{\delta}\frac{\alpha}{x} \left( \sqrt{1+\frac{2x}{\alpha}} - 1  \right ) \left ( 1- x \right ) \right ] \ , \qquad &&
\end{eqnarray}
where we have defined $\alpha:=b/2a$. Finally, introduce notation $\Delta LEB(r|r^{\prime}):=LEB_r(x)-LEB_{r^{\prime}}(x)$ in order to compare each replicator type. Hence, for ${\bf h}$ and ${\bf p}$ against ${\bf s}$ we derive

\begin{eqnarray}\label{DLEB Hyperbolic-Simple}
\Delta LEB({\bf h}|{\bf s})&=&\log \left( \frac{h}{g} x \right ) \ , \\
\label{DLEB Parabolic-Simple}
\Delta LEB({\bf p}|{\bf s})&=&\log \left[  \frac{c}{g}\frac{\alpha}{x} \left( \sqrt{1+\frac{2x}{\alpha}} - 1  \right )   \right ] \ ,
\end{eqnarray}
while, $\Delta LEB({\bf h}|{\bf p})= \Delta LEB({\bf h}|{\bf s}) - \Delta LEB({\bf p}|{\bf s})$. Notice that, since all replicators decay mechanism has been chosen to be equivalent (see Appendix B), then relative bounds $\Delta LEB(r|r^{\prime})$ are $\delta-$independent. Figure 3a-3f show various curves \eqref{DLEB Hyperbolic-Simple} and \eqref{DLEB Parabolic-Simple} against the density value $x$.

\begin{figure}[t]
\centering
\includegraphics[width=15 cm]{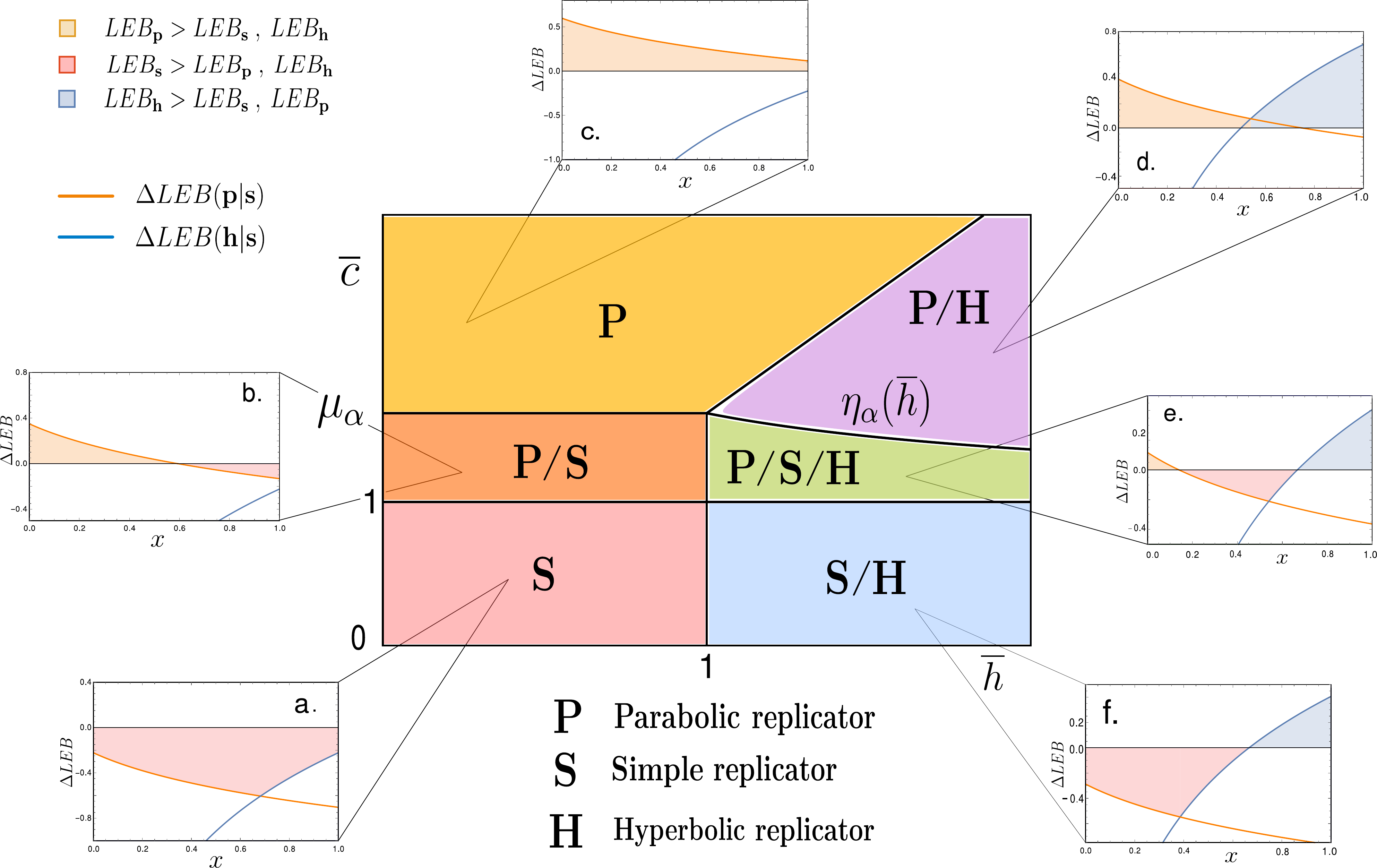}
\vspace{0.25 cm}
\caption{The central diagram corresponds to the space spanning the reduced variables $(\bar{h},\bar{c})$. It is possible to distriguish six phases depending on the dominance of the LEB of each replicator type, $\{\mathbf{S},\mathbf{H},\mathbf{P}\}$. (a) $\mathbf{S}$-dominant (the simple replicator LEB exceeds that of both parabolic and hyperbolic); (b) $\mathbf{P}/\mathbf{S}$ i.e., at low densities, it is $\mathbf{S}$-dominant, while, for $x>x_{ps}$ we observe $\mathbf{S}$ dominance; (c) $\mathbf{P}$-dominant at all density values; (d) $\mathbf{P}/\mathbf{H}$ $\mathbf{P}$ at low densities and $\mathbf{H}$-dominant for $x>x_{ph}$; (e) $\mathbf{P}/\mathbf{S}/\mathbf{H}$ where the three replicators share dominance at some point, jumping orderedly at density values $x_{ps}<x_{sh}$; (f) here simple replicators have a higher LEB at low densities than parabolic ones, but hyperbolic ones take over at high densities, $x>x_{sh}$. Numerical values of $(\bar{h},\bar{c})$ for each plot are: (a) (0.8,0.8); (b) (1.42,0.8), (c) (1.82,0.8), (d)(1.5,2); (e) (1.125,1.5); (f) (0.75,1.5), while $\alpha=0.5$ for all graphs.} 
\end{figure}  

Focusing on the limiting cases where the lower bounds between distinct replicators coincide, $\Delta LEB(r|r^{\prime})=0$, it is possible to derive the density values for which the LEB for replicator $r$ exceeds that of replicator $r^{\prime}$ and viceversa.
This is an interesting exercise since minimal entropy production can provide a guideline for thermodynamically advantatgeous processes. Bare in mind that exploring LEBs does not include the full picture, though, as fluctuations can shift irreversibility as discussed above \cite{Perunov16-PRX}.

Thus, let us define the LEB crossover density $x_{rr^{\prime}}$ from $r$-LEB dominance to $r^{\prime}$-LEB dominance, or, simply, $\Delta LEB(r|r^{\prime})\big{|}_{x_{rr^{\prime}}}=0$. Working with reduced variables $\bar{h}:=h/g$ and $\bar{c}:=c/g$ we derive $x_{rr^{\prime}}=x_{rr^{\prime}}(\bar{h},\bar{c})$ following \eqref{DLEB Hyperbolic-Simple} and \eqref{DLEB Parabolic-Simple}:

\begin{eqnarray}\label{Crossover lines}
x_{sh}=\bar{h}^{-1} \ ,\ \ x_{ps}=2\alpha \bar{c}\left(\bar{c}-1\right) \ ,\ \ x^3_{ph} + \frac{2\alpha \bar{c}}{\bar{h}}\left(x_{ph}-\frac{\bar{c}}{\bar{h}}\right)=0 \ ,
\end{eqnarray}
where the equation for $x_{ph}$, the density value where LEB dominance shifts from parabolic hyperbolic is given in an implicit form\footnote{Algebraic analysis shows that the equation for $x_{ph}$ contains a single real root.}. On the other hand, $0<x_{rr^{\prime}}<1$ must be held, as it stands for a density variable.

This considerations allow for a construction of a diagram $(\bar{h},\bar{c})$ where space is divided into sections characterised by the replicator-types that display a dominant LEB. For instance, for $\bar{h}, \bar{c}<1$ the simple replicator's lower entropic production bound is always larger than the other two types, we denote this sector of the phase space by $\mathbf{S}$ (red shaded region in Figure 3). Most regions, however, will display dominance of entropy production by one type or replicators for a range of densities, and shift dominance over another type for another range of $x$ values (see Figures 3b \& 3d-3f).

The lines separating sections of LEB dominance are given by the following set of inequalities, all derived from the results above:

\begin{eqnarray}
&\mathbf{P}& \Leftrightarrow \left\{\bar{c}>1 \ \ \& \ \ \ 0<\bar{h}\right\} \ , \\
&\mathbf{S}& \Leftrightarrow \left\{\bar{c}<\mu_{\alpha} \ \ \& \ \ \ \bar{h}<1\right\}\cup\left\{\bar{c}<\eta_{\alpha}(\bar{h}) \ \ \& \ \ \ \bar{h}>1\right\} \ , \\
&\mathbf{H}& \Leftrightarrow \left\{\bar{c}<\mu_{\alpha} \ \bar{h}\ \ \&  \ \ \ \bar{h}>1\right\} \ ,
\end{eqnarray}
with the associated functions
\begin{eqnarray}\label{Upper line simple I}
\mu_{\alpha}:=\frac{1}{2}\left(1+\sqrt{1+\frac{2}{\alpha}}\right) \ , \ \ \eta_{\alpha}(\bar{h}):=\frac{1}{2}\left(1+\sqrt{1+\frac{2}{\alpha\bar{h}}}\right) \ .
\end{eqnarray}

Notice that, in several patches of the space of parameters depicted in Figure 3, LEB dominance is dependent on specific density values. Also, $\Delta LEB(r|r^{\prime})$ functions behave such that LEB dominance (if occurs) always appears ordered as $\mathbf{P}$, $\mathbf{S}$ and $\mathbf{H}$, respectively. This ordered sequence can be understood as an indication of a subjacent thermodynamical constrain for these pre-biotic replicating systems. Finally, notice that this analysis has been performed with fixed value of $\alpha$. Nonetheless, shifting the values of this internal parameter does not substantially modifies the structure of the phase space given in Figure 3, in fact, its topological arrangement will remain invariant.

Henceforth, from macroscopical considerations involving both coarse-grained values for the coupling constants $\{g,h,c\}$ and parameter $\alpha$, we are able to derive a phase space compartimentalization such that a classification based on the lower entropy production bounds for each replicator type. As shown, from the subsequent bounds it is infered that the parabolic replicator generates more entropy at replication for low densities while so does the hyperbolic replicator at high $x$ values, leaving the simple replicator inbetween. 

%% If the documentclass option "submit" is chosen, please insert a blank line before and after any math environment (equation and eqnarray environments). This ensures correct linenumbering. The blank line should be removed when the documentclass option is changed to "accept" because the text following an equation should not be a new paragraph. 

%%%%%%%%%%%%%%%%%%%%%%%%%%%%%%%%%%%%%%%%%%

\section{Discussion}

A significant gap in our understanding of evolution, particularly in relation with early events and simple living systems, stems from the lack of a physical theory incorporating a thermodynamic description of replication dynamics. Recent work has addressed this problem revealing a powerful connection between entropy production and transition probability, and pointed to the relevance in biological systems \cite{England13-JChemPhys, Andrieux08-PNAS}. Such connection can be efficiently exploited to analyse, under the coarse-graining described above, the general tendency of a Darwinian replicator to replicate itself. In this way, it is possible in particular to compare the efficiency of different classes of replicators by looking at their relative lower entropy bounds.

Instead of a direct comparison of the systems measurable transition replication rates, this framework focuses on how, via a coarse-graining procedure, these parameters are resulting from the interplay of the many internal degrees of freedom. This technique ultimately leads to the estimation of the lower entropic bounds for each replicator. We interpret these thermodynamical bounds as a consistent way of comparing and evaluating the likelihood of observing different classes of replicators. This is summarised in the phase diagram shown in Figure 3 where the relative dominance of each class is indicated. Notice that the analysis above does not involve competition between the replicator classes. All computations for the entropic bounds are done by considering the replicators to be evolving separately (see Appendix B for details).

Even at this level of description, we can see how the coarse-graining predicts what to expect and thus a physical approach to the constraints operating on the classes of replicators that can emerge through evolution. The diagrams reveal the threshold conditions that would allow particular types of replicator to thrive or coexist in a competing scenario. In some domains only Malthusian dynamics are thermodynamically dominant, while, in others, parabolic replicators seem to be more efficient at generating entropy. Also, in some regions, a combination of parabolic and hyperbolic (cooperator) agents would share dominance. Overall, there is a roboust characterization of dominance related to the density of the system, revealing a preferencial order in going from low to high densities.

Future work should be aimed at the construction of theoretical microscopic models such that coarse-graining operations can be unambiguously defined and subsequent operations may be computed in order to obtain the emergent transition rates. This would yield a deeper understanding of both the coarse-graining operation and how some biological systems seem to be able to operate at the edge of what is possible. Such approach can lead to novel insights into the problem of how major evolutionary transitions occur.

%This section is not mandatory, but can be added to the manuscript if the discussion is unusually long or complex.

%%%%%%%%%%%%%%%%%%%%%%%%%%%%%%%%%%%%%%%%%%
\vspace{6pt}

%%%%%%%%%%%%%%%%%%%%%%%%%%%%%%%%%%%%%%%%%%

%%%%%%%%%%%%%%%%%%%%%%%%%%%%%%%%%%%%%%%%%%
\begin{acknowledgements}
The authors thank the Complex Systems Lab members for fruitful discussions. Special thanks to Gerda Taro for her inspiring ideas. This work was supported by the Botín Foundation by Banco Santander through its Santander Universities Global Division, the Spanish Ministry of Economy and Competitiveness, grant FIS2016-77447-R MINEICO/AEI/FEDER, and the Santa Fe Institute.
\end{acknowledgements}

%%%%%%%%%%%%%%%%%%%%%%%%%%%%%%%%%%%%%%%%%%

\section{Appendix A}

%%%%%%%%%%%%%%%%%%%%%%%%%%%%%%%%%%%%%%%%%%

The argument for the effective kinetical law for the parabolic replicator goes as following: let $y=[AA]$ (concentration of associated molecules), and $z=[A]$ (concentration of dissociated molecules). Thus, define $x=2y+z$ as the gross stechiometric concentration of molecules of type $A$, regardless of configuration. Assuming that the time-scale of the replication reaction (here moduled by ratio $c$) is much larger than the association/dissociation processes, then, by focusing on the dynamics of replication, we can assume balanced equilibrium

\begin{equation}\label{Association/Dissociation Equilibrium}
by=az^2 \Leftrightarrow z=\sqrt{\frac{b}{a}} \ y^{1/2} \ .
\end{equation}

Then, analyzing the dynamics of the replication reaction, which is moduled by the parameter $c$,

\begin{equation}\label{Kinetic Replication A to AA}
\frac{dy}{dt}=cz=c \ \sqrt{\frac{b}{a}} \ y^{1/2} \ ,
\end{equation}

while the kinetics for the gross concentration $x$ is obtained by using \eqref{Association/Dissociation Equilibrium} as

\begin{equation}\label{Kinetic Replication A and AA}
\frac{dx}{dt}=2\frac{dy}{dt}+\frac{dz}{dt}=\frac{cb}{2a}+ 2 c \  \sqrt{\frac{b}{a}} \ y^{1/2} ,
\end{equation}

but, as $a\gg b$, then the equilibrium of the association/dissociation reaction is very much unbalanced in favour of the associated molecular configuration $AA$, which implies that $x\approx 2y$. Thus, we conclude that the kinetics for $x$ is given by

\begin{equation}\label{Parabolic Kinetics I}
\frac{dx}{dt}=\frac{cb}{2a}+  c \  \sqrt{\frac{2b}{a}} \ x^{1/2} .
\end{equation}

Truncating at leading terms in $\left ( b / a \right )$, we derive

\begin{equation}\label{Parabolic Kinetics II}
\frac{dx}{dt}\approx \rho x^{1/2} ,
\end{equation}

with $\rho=c \  \sqrt{\frac{2b}{a}}$.

%%%%%%%%%%%%%%%%%%%%%%%%%%%%%%%%%%%%%%%%%%

\section{Appendix B}

%%%%%%%%%%%%%%%%%%%%%%%%%%%%%%%%%%%%%%%%%%

Consider a well-mixed urn filled with $N$ elements that can be characterized as dead or alive. Notice that such a characterization embodies some kind of coarse-grained measure, since we are deliberately ignoring (integrating) all internal degrees of freedom for each element. Denote by $n<N$ the number of {\em active} (alive) elements in the urn at a given time $t$. In the following sections we will derive the coarse-grained (mesoscopical) time-dependent dynamics. Hence, for each replicator type, let us construct a master equation of the form

\begin{eqnarray}\label{MEQ General Appendix}
\frac{dP(n,t)}{dt}= \sum_{m\neq n} \omega\left( n| m \right) P(m,t) - \sum_{m\neq n} \omega \left ( m | n \right) P(n,t) \ , 
\end{eqnarray}

while restricting the dynamics to a first-step process and introducing a natural (single-particle) decay process moduled by parameter $\delta$ that will be equivalent to all replicating motifs.

\subsection{Appendix B.1}

Beginning with the simple replicator, introduce the following rules (see Figure A1):

\begin{enumerate}
\item	Pick an element of the urn at random.
\item	If active, with probability $g$, pick a second element at random and (if not active) activate.
\item	Pick an element at random again.
\item	If active, with probability $\delta$, deactivate.
\end{enumerate}

\begin{figure}[t]
\centering
\includegraphics[width=16 cm]{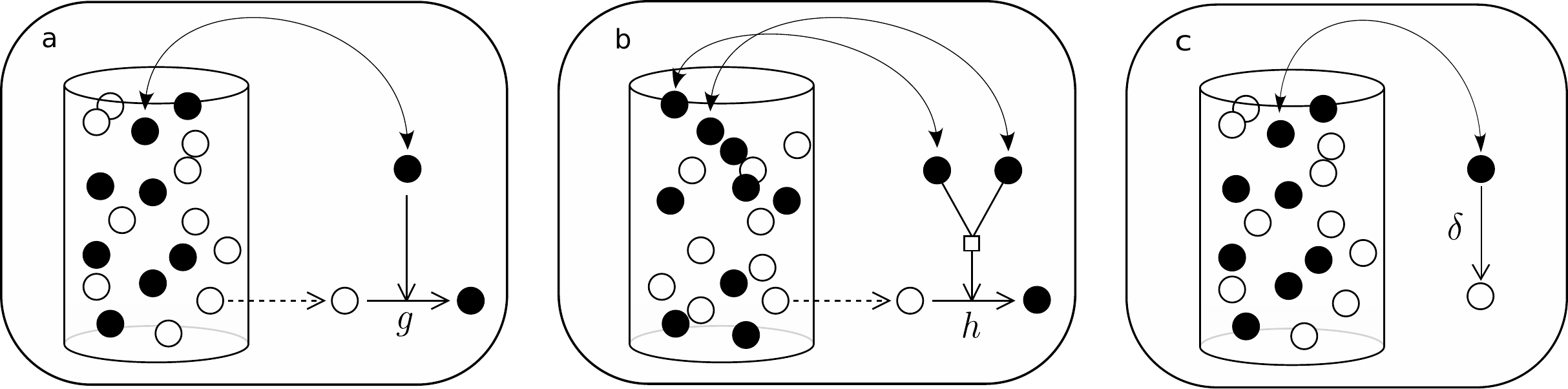}
\caption{A schematization of the rules of replication in an urn model. Active chains are drawn as filled balls and inactive chains are white balls. (a) represents the action of selecting an active chain replicating following the simple replicator mechanism. (b) shows the replicating process of a hyperciclic replicator. (c) corresponds to the decay which, for the purpose of this work, is supposed to act equivalently in each replicator-type.}
\end{figure} 

For the simple replicator, using the notation on Table 1, plus adding a single-particle decay process, 

\begin{eqnarray}\label{Simple Transition Rules}
\omega ( n | n-1 ) = \left ( \frac{n-1}{N} \right ) \left ( \frac{N-n+2}{N} \right )g \ \ \ ; \ &
\omega ( n | n+1 ) = \left ( \frac{n+1}{N} \right ) \delta
\end{eqnarray}

Then, for sufficiently large $N$, it is possible to approach the dynamics by

\begin{eqnarray}\label{MEQ Simple II}
\frac{dP(n,t)}{dt} = g\left ( \frac{n}{N} \right ) \left ( 1 - \frac{n}{N} \right ) \left [ P(n-1,t)-P(n,t) \right ]  - \delta \left ( \frac{n}{N} \right ) \left [ P(n,t)-P(n+1,t) \right ] \ , 
\end{eqnarray}

\subsection{Appendix B.2}

Consider now the dynamics of hyperbolic replicators. Following the rules summarized in Table 1,
introduce the algorithm (see Figure A1):

\begin{enumerate}
\item	Pick an element of the urn at random.
\item	If active, pick a second element at random.
\item	If active, with probability $h$, pick a third element at random and (if not active) activate.
\item	Pick an element at random again.
\item	If active, with probability $\delta$, deactivate.
\end{enumerate}

Hence, restricting the dynamics to a first-step process, we may deduce the following transition probabilities

\begin{eqnarray}\label{Hyperbolic Transition Rules} 
\omega ( n | n-1 ) = \left ( \frac{n-1}{N} \right ) \left ( \frac{n-2}{N} \right ) \left ( \frac{N-n+3}{N} \right )h \ \ \ ; \ &
\omega ( n | n+1 ) = \left ( \frac{n+1}{N} \right ) \delta 
\end{eqnarray}

which, for $N\gg 1$, lead to the master equation

\begin{eqnarray}\label{MEQ Hyperbolic II}
\frac{dP(n,t)}{dt} = h\left ( \frac{n}{N} \right )^ 2 \left ( 1 - \frac{n}{N} \right ) \left [ P(n-1,t)-P(n,t) \right ]  - \delta \left ( \frac{n}{N} \right ) \left [ P(n,t)-P(n+1,t) \right ] \ , 
\end{eqnarray}

\subsection{Appendix B.3}

Finally, let us derive the macroscopical dynamics for a parabolic replicator by implementing the following set of rules on an urn of $N$ elements (see Figure A2):

\begin{enumerate}
\item	Pick an element of the urn at random. If active, then: {\em (i)} if in associated state ($AA$) then, with probability $a$, dissociate and iterate. {\em (ii)} if dissociated, pick a second element and, if active, with probability $b$, associate. Iterate this process until equilibrium is reached for association/dissociation reaction.
\item	Pick an element of the urn at random. If active, pick a second element at random, if empty, with probability $c$, replicate.
\item	Pick an element of the urn at random. If active, with probability $\delta$, deactivate.
\end{enumerate}

\begin{figure}[t]
\centering
\includegraphics[width=15 cm]{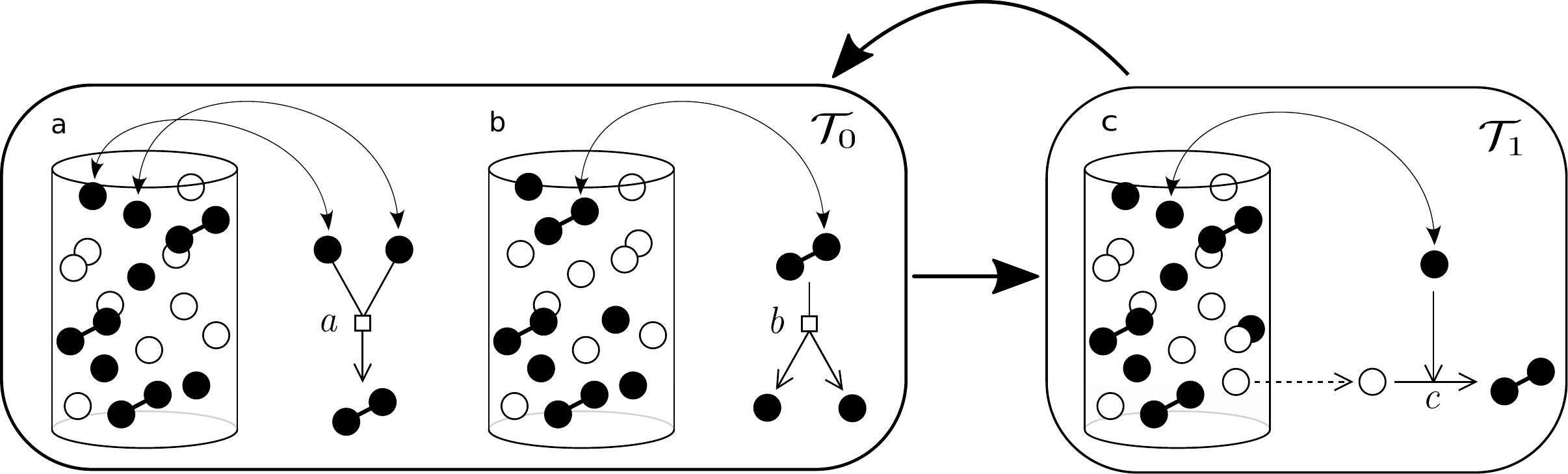}
\caption{This diagram shows how the urn model of parabolic replicators is implemented. (a) and (b) correspond to the rapid association/dissociation reactions, which are supposed to equilibrate in much shorter time-scales than the replicating process, which is shown in (c), i.e., $\tau_0\ll\tau_1$. The process of equilibration (left box) is iterated a large number of times before the loop goes into the replication process (right box).}
\end{figure} 

The situation for the parabolic replicator is a peculiar one, for one thing, it involves two characteristical time-scales, a rapid one, concerning the association/dissociation process (see Appendix A), and the replication process. In order to approximate the transition rates let us define $k$ as the number of associated pairs, $AA$, and $m$ as the number of dissociated active elements in the urn, $A$. Let $N$ be the total number of elements in the urn, including associated, dissociated and deactivated elements. Let us denote by $n$ the total number of active elements, regardless of configuration, then, $n=2k+m$. Now, assuming rapid equilibration of the association/dissociation reaction in \eqref{Parabolic Reactions},

\begin{eqnarray}\label{Balanced Parabolic Urn}
\left ( \frac{2k}{N} \right ) b = \left ( \frac{m}{N} \right )^2 a \Leftrightarrow m^2=\frac{2b}{a}kN \ ,
\end{eqnarray}

which can be related to the number $n$ by

\begin{eqnarray}\label{Relation between m and n}
m^2=\frac{b}{a}(n-m)N \Leftrightarrow m(n)=\frac{bN}{2a} \left( \sqrt{1+\frac{4an}{bN}} - 1  \right ) \ ,
\end{eqnarray}

where we neglect the negative root, as it is non-physical. Hence, it is now possible to construct a master equation for the first-step process of replication as in \eqref{Parabolic Reactions}, with

\begin{eqnarray}\label{Parabolic Transition Rules} 
\omega ( n | n-1 ) = \left ( \frac{m+1}{N} \right ) \left ( \frac{N-n+1}{N} \right )c \ \ \ ; \ &
\omega ( n | n+1 ) = \left ( \frac{n+1}{N} \right ) \delta
\end{eqnarray}

which, for $N\gg 1$, and using \eqref{Relation between m and n} lead to

\begin{eqnarray}\label{MEQ Parabolic II}
\frac{dP(n,t)}{dt} &=& \frac{bc}{2a} \left( \sqrt{1+\frac{4an}{bN}} - 1  \right ) \left ( 1 - \frac{n}{N} \right ) \left [ P(n-1,t)-P(n,t) \right ] \nonumber \\
&-& \delta \left ( \frac{n}{N} \right ) \left [ P(n,t)-P(n+1,t) \right ] \ , 
\end{eqnarray}

\end{document}